\documentclass[superscriptaddress,preprint]{revtex4}
\usepackage{mathrsfs}
\usepackage{amssymb}
\usepackage[tbtags]{amsmath}
\usepackage{graphicx}
\usepackage{epsfig,graphicx,times}
\usepackage{color}
\usepackage{subfigure}

\begin{document}
\title{Weak values could reveal the hidden effects of quantum interactions}
\author{Miao Zhang\footnote{miaozhangphys@gmail.com}}
\affiliation{Department of Physics, Southwest Jiaotong University,
Chengdu 610031, China}
\date{\today}

\begin{abstract}
Due to the reduced probability of successful post-selection, the
weak-value amplification seems to be unavailable for the
parameter-estimation. Here, we show theoretically that, some effects
due to the weak interactions present only in the properly
post-selected sub-ensemble, however are canceled by themselves in
the total ensemble. From this point of view, the post-selection
induced weak value could be one of the feasible methods for
measuring the weak interaction, since the standard measurement does
not work. Additionally, we employ the system of trapped ions to
simulate the weak measurement and calculate relevant results without
the frequently-used weak interaction approximation.
\end{abstract}
\maketitle

\section{Introduction}
The conception of quantum weak measurements was noticed by Aharonov,
Albert, and Vaidman (AAV) in 1988~\cite{AAV}. The weak measurements
have two key features: the couplings between the quantum systems are
sufficiently weak, and the relevant observable quantities (act as
the pointers) are measured in the properly post-selected
sub-ensemble. The outcome of weak measurements is the so-called weak
value
\begin{equation}
A_w=\frac{\langle \psi_f|\hat{A}|\psi_i\rangle}{\langle
\psi_f|\psi_i\rangle}\,,
\end{equation}
where $|\psi_i\rangle$, $|\psi_f\rangle$, and $\hat{A}$ are the
initial state, post-selected state, and the observable operator of
the measured system, respectively. Comparing to the usual quantum
measurement $A=\langle\psi_i|\hat{A}|\psi_i\rangle$, the weak value
$A_w$ could generate some interesting
results~\cite{AAV,Nature-wavefuction,Science-trajectory,Hall
Effect,Anomalous}, as the post-selection $|\psi_f\rangle$ is
induced.

Recently, the weak value has been employed to study the foundational
questions of quantum mechanics, e.g., the Hardy's paradox~\cite{HD},
the Leggett-Garg inequality~\cite{LG}, the Heisenberg's uncertainty
relation~\cite{Uncertain}, the Cheshire Cat~\cite{Cat}, and the
superluminal velocities~\cite{Superluminal1}. Much attentions are
also paid on the practical applications of the weak value, e.g.,
detecting tiny values of a quantity~\cite{RMP}. Considering an
observable quantity depends on $g|A_w|$ (with $g$ being a small
parameter)~\cite{Jozsa}, then the weak value $A_w$ can significantly
change the pointer with $|\langle
\psi_f|\psi_i\rangle|\rightarrow0$, or say amplifying the weak
signal of
$g$~\cite{Amplification0,Amplification1,Amplification2,Amplification3,Amplification4,Amplification-arXiv,arXiv}.
However, serval recent papers claim this amplification offering no
fundamental metrological advantage, due to the necessarily reduced
probability of successful
post-selection~\cite{SNR,SSNR,Fisher1,Fisher2}, i.e., $|\langle
\psi_f|\psi_i\rangle|^2\rightarrow0$.

In this work, we recommend an useful weak value, which could reveal
the hidden effects of the weak quantum interactions. Here the hidden
effects mean that, the $g$-induced changes of some observable
quantities can not be detected by the usual quantum measurement
(with the total ensemble). However, these changes can be
conditionally measured by the weak value technique. If these special
observable quantities are the better ones for experimentally
detecting, then the weak value technique could be considered as an
effective candidate for measuring the small parameter $g$ of weak
interaction. Moreover, we prove the relevant results by a simple
example which is calculated without the frequently-used weak
interaction approximation, and employ the powerful
experimental-system of trapped ions to simulate the weak
measurements.

This work is organized as follows. In Sec. II, we show that the
usual measurement can not detect some special effects of the weak
quantum interactions. In Sec. III, we show the conditional
measurements could reveal these hidden effects, and calculate an
example without the usual weak interaction approximation. In Sec.
IV, we employ the trapped ions to simulate the weak measurement, and
finally give a conclusion in Sec. V.

\section{The vanished effects of one-order $g$}
We consider the standard von Neumann measurement Hamiltonian
\begin{equation}
\hat{H}=\hbar g_0\hat{A}\hat{P}\,.
\end{equation}
Where, $\hbar$ is the Planck constant divided by $2\pi$, $\hat{A}$
and $\hat{P}$ are respectively the operators of the qubit and the
pointer (with the coupling frequency $g_0$ between them). After the
above interaction, the final state of the total system can be
formally written as
\begin{equation}
|\psi_f\rangle=e^{-ig\hat{A}\hat{P}}|S_0\rangle|\phi_0\rangle\,.
\end{equation}
Where, $g=g_0t$ (with the interaction duration $t$) describes the
coupling strength between the pointer and the qubit, $|S_0\rangle$
and $|\phi_0\rangle$ are respectively the initial states of the
qubit and the pointer. Considering the interaction is sufficiently
weak, i.e., $g\rightarrow0$, then the final state (3) can be
approximately written as
\begin{equation}
|\psi_f\rangle\approx|S_0\rangle|\phi_0\rangle
-ig\hat{A}\hat{P}|S_0\rangle|\phi_0\rangle\,.
\end{equation}
by neglecting the high order of $O(g^2)$.

Directly, the probability of finding the pointer state $|x\rangle$
reads
\begin{equation}
\begin{array}{l}
I=\left|\langle x|\psi_f\rangle\right|^2
\\
\\
\,\,\,\,\,\approx\left|\langle
x|\phi_0\rangle\right|^2-ig\left(\langle \phi_0|x\rangle\langle
x|\hat{P}|\phi_0\rangle\langle
S_0|\hat{A}|S_0\rangle-\text{c.c}\right)
\\
\\
\,\,\,\,\,=I_0\left[1-ig(P_w\langle
S_0|\hat{A}|S_0\rangle-\text{c.c})\right]
\\
\\
\,\,\,\,\,=I_0\left[1+2g\text{Im}(P_w\langle
S_0|\hat{A}|S_0\rangle)\right]
\end{array}
\end{equation}
with $I_0=\left|\langle x|\phi_0\rangle\right|^2$ and
\begin{equation}
P_w=\frac{\langle x|\hat{P}|\phi_0\rangle}{\langle
x|\phi_0\rangle}\,.
\end{equation}
Above, $\text{c.c}$ is the conjugate complex number of the first
term in the bracket, and the high order of $O(g^2)$ has been
neglected again. This approximation shows that: when
\begin{equation}
P_w\langle S_0|\hat{A}|S_0\rangle=\text{Real}\,,
\end{equation}
there is no effect due to the one-order $g$. Thus, we say that: the
observable $I$ is insensitive to the parameter $g$, and the
weak-value amplification may be useful (when $I$ is favorable for
experimentally detecting).

Worth of note that, if $|x\rangle$ is the state of atoms at position
$x$, the observable $I$ can be regarded as the {\it``lightness"} of
atoms beam. This is similar to the light intensity, which is
proportional to $|\vec{E}(x)|^2$ of the Maxwell field $\vec{E}(x)$.
Specially, we consider the two-level transition operator
$\hat{A}=\hat{\sigma}_x=|e\rangle\langle g|+|g\rangle\langle e|$ and
the initial state $|S_0\rangle=\alpha|g\rangle+\beta|e\rangle$ of
the qubit. Here, $\alpha$ and $\beta$ are the normalized
coefficients of the two orthogonal states $|g\rangle$ and
$|e\rangle$. When $P_w=\text{Real}$, the Eq.~(5) can be rewritten as
\begin{equation}
\begin{array}{l}
I=I_0\left[1-igP_w\left(\alpha^*\beta+\beta^*\alpha-\text{c.c}\right)\right]\\
\,\,\,\,\,=I_0\left[1+2gP_w\text{Im}(\alpha^*\beta)-2gP_w\text{Im}(\alpha^*\beta)\right]\\
\,\,\,\,\,=I_0+I_g-I_g\\
\,\,\,\,\,=I_0
\end{array}
\end{equation}
with $I_g=2gP_w\text{Im}(\alpha^*\beta)I_0$. This result shows that,
under the condition of (7), the $g$-induced effects are accurately
canceled by themselves.

\section{The weak measurements}
With the post-selected state $|S_f\rangle$ of the qubit, the
probability of finding the pointer state $|x\rangle$ reads:
\begin{equation}
\begin{array}{l}
I_{s}=\left|\langle x|\langle S_f|\psi_f\rangle\right|^2\\
\\
\,\,\,\,\,\,\approx\left|\langle S_f|S_0\rangle\langle
x|\phi_0\rangle -ig\langle S_f|\hat{A}|S_0\rangle\langle
x|\hat{P}|\phi_0\rangle\right|^2\\
\\
\,\,\,\,\,\,\approx I_0\times\left|\langle
S_f|S_0\rangle\right|^2\times\left[1-igP_w(A_w-\text{c.c})\right]
\\
\\
\,\,\,\,\,\,=I_0\times\left|\langle
S_f|S_0\rangle\right|^2\times\left[1+2gP_w\text{Im}(A_w)\right]
\end{array}
\end{equation}
with the weak value
\begin{equation}
A_w=\frac{\langle S_f|\hat{A}|S_0\rangle}{\langle S_f|S_0\rangle}\,.
\end{equation}
The Eq.~(9), with $A_w\neq\text{Real}$, means that the effect due to
the one-order $g$ presents now in the post-selected sub-ensemble of
$|S_f\rangle$. On the other hand, within the remaining sub-ensemble
of $|S^f\rangle$, we have the similar result: the probability of
finding $|x\rangle$ reads
\begin{equation}
I^{s}=I_0\times\left|\langle
S^f|S_0\rangle\right|^2\times\left[1+2gP_w\text{Im}(A^w)\right]
\end{equation}
with the corresponding weak value
\begin{equation}
A^w=\frac{\langle S^f|\hat{A}|S_0\rangle}{\langle S^f|S_0\rangle}\,.
\end{equation}

Considering $|S_f\rangle=|g\rangle$ and $|S^f\rangle=|e\rangle$, the
Eq.~(9) and (11) can be further written as
\begin{equation}
I_{s}=I_0\times \left[|\alpha|^2+2gP_w\text{Im}(\alpha^*\beta)
\right]\,,
\end{equation}
\begin{equation}
I^{s}=I_0\times \left[|\beta|^2-2gP_w\text{Im}(\alpha^*\beta)\right]
\end{equation}
and then the total probability of finding the pointer state
$|x\rangle$ reads
\begin{equation}
\begin{array}{l}
I_{s}+I^{s}=I_0\times
\left[1+2gP_w\text{Im}(\alpha^*\beta)-2gP_w\text{Im}(\alpha^*\beta)\right]\\
\,\,\,\,\,\,\,\,\,\, \,\,\,\,\,\,\,\,\,=I_0+I_g-I_g
\\
\,\,\,\,\,\,\,\,\,\, \,\,\,\,\,\,\,\,\,=I_0
\end{array}
\end{equation}
This result is same to that of Eq.~(8): the $g$-induced effects
vanished from the usual measurement with the total ensemble.
However, the Eq.~(13) and (14) show that, the $g$-induced changing
of observable $I_0$ presents magically in the properly post-selected
sub-ensemble (see the dramatic Fig.~1). Moreover, it is easily to
calculate the maximal $\text{Im}(\alpha^*\beta)=1/2$ (according to
the normalized condition $|\alpha|^2+|\beta|^2=1$). Thus, the
Eq.~(13) and (14) reduce further to
$I_{s}=I_0\times(\frac{1}{2}+gP_w)$ and
$I^{s}=I_0\times(\frac{1}{2}-gP_w)$, respectively.
\begin{figure}[tbp]
\includegraphics[width=10cm]{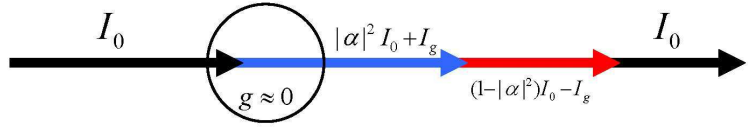}
\caption{(Color online) Sketch for the observable $I_0=\left|\langle
x|\phi_0\rangle\right|^2$ affected by the one-order $g$. Under the
situation of Eq.~(7), there is no effect of $g$ with the usual
measurement (the dark line). However, in the properly post-selected
sub-ensemble (the red or blue line), the $g$-induced effect (i.e.,
$I_g$) appears.}
\end{figure}

We now calculate an example without the weak interaction
approximation preformed in Eq.~(4). Considering
$\hat{P}=\hat{a}^\dagger\hat{a}$ and $\hat{A}=\hat{\sigma}_x$, the
Hamiltonian (2) reads
\begin{equation}
\hat{H}=\hbar g_0\hat{a}^\dagger\hat{a}\hat{\sigma}_x\,,
\end{equation}
with $\hat{a}^\dagger$ and $\hat{a}$ being respectively the creation
and annihilation operators in the Fock state
presentation~\cite{Fock}. Without of generality, we write the
initial state of the pointer as
$|\phi_0\rangle=\sum_{n}c_n|n\rangle$, i.e., a superposed Fock state
with the occupancy number $n\geq0$ and the normalized coefficients
$\sum_n|c_n|^2=1$. Consequently, the initial state of the total
system reads
\begin{equation}
|\psi_0\rangle=|\phi_0\rangle(\alpha|g\rangle+\beta|e\rangle)
=\sum_{n}c_n(\alpha|g\rangle+\beta|e\rangle)|n\rangle
\end{equation}
and the final state reads
\begin{equation}
\begin{array}{l}
|\psi_f\rangle=e^{-ig\hat{a}^\dagger\hat{a}\hat{\sigma}_x}|\psi_0\rangle
=\sum_{n} c_n(\alpha
e^{-ig\hat{a}^\dagger\hat{a}\hat{\sigma}_x}|g\rangle|n\rangle+\beta
e^{-ig\hat{a}^\dagger\hat{a}\hat{\sigma}_x}|e\rangle|n\rangle)
\\
\,\,\,\,\,\,\,\,\,\,\,\,= \sum_{n} c_n\left(\alpha
e^{-ign\hat{\sigma}_x}|g\rangle+\beta
e^{-ign\hat{\sigma}_x}|e\rangle\right)|n\rangle
\\
\,\,\,\,\,\,\,\,\,\,\,\,=\sum_{n} c_n
\left\{\alpha[\cos(gn)|g\rangle-i\sin(gn)|e\rangle]+\beta[\cos(gn)|e\rangle-i\sin(gn)|g\rangle]\right\}|n\rangle
\\
\,\,\,\,\,\,\,\,\,\,\,\,=\sum_{n} c_n
\left(\eta_{gn}|g\rangle+\eta_{en}|e\rangle\right)|n\rangle
\end{array}
\end{equation}
with $\eta_{gn}=\alpha\cos(gn)-i\beta\sin(gn)$ and
$\eta_{en}=\beta\cos(gn)-i\alpha\sin(gn)$. Immediately, we have
\begin{eqnarray}
\left\{
\begin{array}{l}
\left|\eta_{gn}\right|^2=|\alpha|^2\cos^2(gn)+|\beta|^2\sin^2(gn)-i\cos(gn)\sin(gn)(\beta\alpha^*-\alpha\beta^*)
\\
\left|\eta_{en}\right|^2=|\beta|^2\cos^2(gn)+|\alpha|^2\sin^2(gn)+i\cos(gn)\sin(gn)(\beta\alpha^*-\alpha\beta^*)
\\
\left|\eta_{gn}\right|^2+\left|\eta_{en}\right|^2=1
\end{array}
\right.
\end{eqnarray}
and the following results.

(I) For the usual measurement of Fock state $|m\rangle$, we have
\begin{equation}
\left|\langle
m|\psi_f\rangle\right|^2=|c_m|^2\left|(\eta_{gm}|g\rangle+\eta_{em}|e\rangle)\right|^2
=I_m
\end{equation}
with $|c_m|^2=I_m$.

(II) For the conditional measurement, we have
\begin{equation}
\left|\langle m|\langle
g|\psi_f\rangle\right|^2=I_m\left|\eta_{gm}\right|^2 \approx
I_m[|\alpha|^2+2gm\text{Im}(\alpha^*\beta)]\,,
\end{equation}
\begin{equation}
\left|\langle m|\langle
e|\psi_f\rangle\right|^2=I_m\left|\eta_{em}\right|^2 \approx
I_m[|\beta|^2-2gm\text{Im}(\alpha^*\beta)]\,,
\end{equation}
Here, we consider $gm\rightarrow0$ and neglect the high order of
$O(g^2)$. Thus, the results are same to that of Eq.~(13) and (14)
with $P_w=m$, and $\left|\langle m|\langle
g|\psi_f\rangle\right|^2+\left|\langle m|\langle
e|\psi_f\rangle\right|^2=\left|\langle
m|\psi_f\rangle\right|^2=I_m$.

\section{Simulating weak measurement in the ion-trap}
\begin{figure}[tbp]
\includegraphics[width=8cm]{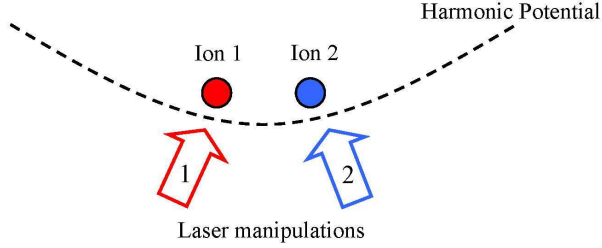}
\caption{(Color online) Sketch for the considered model to implement
the weak measurement with trapped ions. Here, two ions are confined
in a single harmonic potential and manipulated by the laser pulses.
Two internal levels of ion $1$ act as the quantum system (qubit),
the center-of-mass (CM) motion of the two ions serves as the
pointer. The laser $1$ is applied to realize the desirable weak
interaction (16) between the qubit and pointer, and the laser $2$
couples the pointer (i.e., the CM mode) to the atomic levels of ion
$2$ for final readout (by the electron shelving
method)~\cite{Ion-RMP}.}
\end{figure}
To simulate the above weak measurement, we consider two ions trapped
in a linear Paul
trap~\cite{Ion-Cirac,Ion-Monroe,Ion-Meekhof,Ion-RMP}, as it showing
dramatically in Fig.~2. The two internal atomic states of ion $1$ is
encoded as the qubit. By the so-called sideband
excitation~\cite{Ion-Monroe,Ion-Meekhof}, this qubit can be coupled
to the center-of-mass (CM) motion of the two vibrational ions, and
consequently the desirable Hamiltonian of (16) could be realized.
The ion $2$, manipulated by the lase beam $2$, is used to
experimentally measure the CM mode. Under the Lamb-Dicke
approximation and the rotating wave approximation, the
first-red-sideband excitation (generated by the laser pulse $1$) is
described by the well-known Hamiltonian~\cite{Ion-RMP}
\begin{equation}
\hat{H}_i=\hbar\Omega_0 \left(e^{i\delta t}\hat{a}\hat{\tau}_+
+e^{-i\delta t}\hat{a}^\dagger\hat{\tau}_-\right)\,.
\end{equation}
Here, $\hat{a}^\dagger$ and $\hat{a}$ are respectively the boson
creation and annihilation operators of the CM mode,
$\hat{\tau}_+=|\uparrow\rangle\langle \downarrow|$ and
$\hat{\tau}_-=|\downarrow\rangle\langle \uparrow|$ are the two-level
flip operators of ion $1$ (with the internal states
$|\downarrow\rangle$ and $|\uparrow\rangle$), and $\Omega_0$ is the
Rabi frequency of the sideband coupling, and $\delta$ is the
relating detuning (decided by the frequencies of the applied laser
beams). When $\delta=0$, the Hamiltonian (23) describes a standard
Jaynes-Cummings (JC) coupling between the internal and external
states of trapped ions.

Under the large detuning: $\Omega\ll\delta$, the time-evolution
operator of Hamiltonian (23) can be approximately written as:
\begin{equation}
\begin{array}{l}
\hat{U}(t)=1+\left(\frac{-i}{\hbar}\right)\int_0^t\hat{H}_i(t_1)dt_1
+\left(\frac{-i}{\hbar}\right)^2\int_0^t\hat{H}_i(t_1) \int_0^{t_1}
\hat{H}_i(t_2)dt_2dt_1+\cdots\approx
e^{-\frac{i}{\hbar}\hat{H}_et}\,,
\end{array}
\end{equation}
with the effective Hamiltonian
$\hat{H}_e=(\hbar\Omega_0^2/\delta)[\hat{a}^\dagger\hat{a}(|\uparrow\rangle\langle
\uparrow|-|\downarrow\rangle\langle
\downarrow|)+|\uparrow\rangle\langle \uparrow|]$~\cite{Our}. In the
interaction picture defined by $\hat{U}_R=\exp(-ig_0
t|\uparrow\rangle\langle \uparrow|)$, this Hamiltonian can be
further written as
\begin{equation}
\hat{H}_e'=\hbar g_0\hat{a}^\dagger\hat{a}\hat{\tau}_z
\end{equation}
with $\hat{\tau}_z=|\uparrow\rangle\langle
\uparrow|-|\downarrow\rangle\langle \downarrow|$ and
$g_0=\Omega_0^2/\delta$.
We define
$|e\rangle=\frac{1}{\sqrt{2}}(|\uparrow\rangle+|\downarrow\rangle)$
and
$|g\rangle=\frac{1}{\sqrt{2}}(|\uparrow\rangle-|\downarrow\rangle)$
being the qubit of ion $1$. These states satisfy the orthogonal
condition $\langle e|g\rangle=0$ and the normalized condition
$\langle e|e\rangle=\langle g|g\rangle=1$. As
$\hat{\tau}_z|e\rangle=|g\rangle=\hat{\sigma}_x|e\rangle$ and
$\hat{\tau}_z|g\rangle=|e\rangle=\hat{\sigma}_x|g\rangle$, the
Hamiltonian (25) can be written as the desirable form $\hat{H}=\hbar
g_0\hat{a}^\dagger\hat{a}\hat{\sigma}_x$ of Eq.~(16) and
consequently get the same results to (18).

Experimentally, the internal atomic states of the trapped ions can
be well measured by serval methods such as the electron shelving
technique~\cite{Ion-RMP}. This method can be briefly described as
follows. Applying a strong laser resonantly drives the transition
between the state $|\uparrow\rangle$ and an auxiliary atomic state
$|\text{aux}\rangle$. If the ion is in state $|\uparrow\rangle$, the
laser-induced resonance fluorescence can be quickly detected,
whereas there is no signal as the transition
$|\downarrow\rangle\Leftrightarrow|\text{aux}\rangle$ is large
detuning. Here, the post-selection (i.e., the detection of state
$|e\rangle$ or $|g\rangle$) needs additional operations, since these
states are the superposition ones of $|\uparrow\rangle$ and
$|\downarrow\rangle$. We apply a single-qubit operation to the ion
$1$, i.e., $\hat{U}_s=\exp(-i\hat{H}_st/\hbar)$ with the Hamiltonian
$\hat{H}_s=\hbar\Omega_s(e^{-i\theta}\hat{\sigma}_++e^{i\theta}\hat{\sigma}_-)$.
By setting the parameters $\Omega_s t=\pi/4$ and $\theta=-\pi/2$, we
have $\hat{U}_s|e\rangle=|\uparrow\rangle$ and
$\hat{U}_s|g\rangle=-|\downarrow\rangle$. Consequently, we have
\begin{equation}
|\psi_f'\rangle=\hat{U}_s|\psi_f\rangle=\sum_{n} c_n
\left(\eta_{gn}\hat{U}_s|g\rangle+\eta_{en}\hat{U}_s|e\rangle\right)|n\rangle=
\sum_{n} c_n
\left(\eta_{en}|\uparrow\rangle-\eta_{gn}|\downarrow\rangle\right)|n\rangle
\end{equation}
and the same results to (20) $\sim$ (22), i.e.,
\begin{eqnarray}
\left\{
\begin{array}{l}
\left|\langle m|\psi_f'\rangle\right|^2=I_m\,,
\\
\left|\langle m|\langle \downarrow|\psi'_f\rangle\right|^2\approx
I_m[|\alpha|^2+2gm\text{Im}(\alpha^*\beta)]\,,
\\
\left|\langle m|\langle \uparrow|\psi'_f\rangle\right|^2\approx
I_m[|\beta|^2-2gm\text{Im}(\alpha^*\beta)]\,.
\end{array}
\right.
\end{eqnarray}

As that demonstrated in many experiments, the external motional
states of trapped ions can be mapped to their internal states by the
laser manipulations~\cite{Ion-Monroe,Ion-Meekhof,Ion-RMP}, and
measured consequently by the electron shelving technique. For
simplicity, we suppose that the CM mode is prepared initially in the
superposition state
$|\phi_0\rangle=\sum_{n=0}^{1}c_n|n\rangle=c_0|0\rangle+c_1|1\rangle$
(with the normalized coefficients $|c_0|^2+|c_1|^2=1$), and the
internal state of ion $2$ is initially in $|\downarrow_r\rangle$. We
apply a first-red-sideband laser pulse to implement the operation
$\hat{U}_r=\exp(-i\hat{H}_rt/\hbar)$ on ion $2$. Here,
$\hat{H}_r=\hbar\Omega_r \left(\hat{a}\hat{\tau}_{2+}
+\hat{a}^\dagger\hat{\tau}_{2-}\right)$ is the JC coupling between
the CM mode and the internal states of ion $2$, with the two-level
flip operators $\hat{\tau}_{2+}=|\uparrow_r\rangle\langle
\downarrow_r|$ and $\hat{\tau}_{2-}=|\downarrow_r\rangle\langle
\uparrow_r|$. When $\Omega_rt=\pi/2$, this operation generates the
logic:
$|0\rangle|\downarrow_r\rangle\rightarrow|0\rangle|\downarrow_r\rangle$,
$|1\rangle|\downarrow_r\rangle\rightarrow(-i)|0\rangle|\uparrow_r\rangle$.
Consequently, we have
\begin{eqnarray}
\left\{
\begin{array}{l}
\left|\langle \uparrow_r|\psi_f''\rangle\right|^2=\left|\langle
m|\psi_f'\rangle\right|^2=I_m\,,
\\
\left|\langle \uparrow_r|\langle
\downarrow|\psi''_f\rangle\right|^2=\left|\langle m|\langle
\downarrow|\psi'_f\rangle\right|^2\approx
I_m[|\alpha|^2+2gm\text{Im}(\alpha^*\beta)]\,,
\\
\left|\langle\uparrow_r|\langle
\uparrow|\psi''_f\rangle\right|^2=\left|\langle m|\langle
\uparrow|\psi'_f\rangle\right|^2\approx
I_m[|\beta|^2-2gm\text{Im}(\alpha^*\beta)]\,.
\end{array}
\right.
\end{eqnarray}
with $|\psi_f''\rangle=\hat{U}_r|\psi_f'\rangle|\downarrow_r\rangle$
and  $m=1$. These results are same to that of Eq.~(27). Worth of
note that, the above operations (i.e., $\hat{U}_s$, $\hat{U}_r$, and
the preparation and readout of the initial and final states) are
that frequently used in the system of laser-manipulated trapped
ions. Thus, the above simulation of weak measurements could be
experimentally feasible. Indeed, the quantum entanglement with eight
ions has been successfully demonstrated (need more operations) in
this powerful experimental platform~\cite{8Ion}.

\section{Conclusion}
In this study, we present a very simple version to explain the
usefulness of weak values. We showed that the weak values work well
under the certain conditions, e.g., Eq.~(7). With this equation,
some effects of the weak interactions present only in the properly
post-selected sub-ensemble, however are canceled by themselves in
the total ensemble. From this point of view, the weak value can be
regarded as one of the effective methods for measuring the weak
interaction, as the usual measurement does not work. We presented an
example to exactly calculate the relevant results. These results
prove that the frequently-used weak interaction approximation is
effective indeed. Additionally, we showed that the famous
experimental-system of trapped ions could also be utilized to
simulate the weak measurements. Finally, we hope this paper could
help understanding the weak values and moving the relevant
researches forward.

{\bf Acknowledgements}: This work was supported by the National
Natural Science Foundation of China Grants No. 11204249.
\\

\end{document}